\documentclass{webofc}
\usepackage[varg]{txfonts}   
\usepackage{listings}
\usepackage{hyperref}
\pdfoutput=1
\begin{document}
\title{Pulsating White Dwarfs}
\author{\firstname{S. O.} \lastname{Kepler}\inst{1}\fnsep\thanks{\href{mailto:kepler@if.ufrgs.br}{\tt kepler@if.ufrgs.br}} \and
        \firstname{Alejandra D.} \lastname{Romero}\inst{1}\fnsep\thanks{\href{mailto:alejandra.romero@ufrgs.br}{\tt alejandra.romero@ufrgs.br}} 
}

\institute{Departamento de Astronomia, Instituto de F\'{\i}sica, Universidade Federal do Rio Grande do Sul, 91501-970 Porto Alegre, RS - Brasil}

\abstract{%
The Sloan Digital Sky Survey has allowed us to increase the number of known white dwarfs by a factor of five and consequently the number of known pulsating white dwarfs also by a factor of five. It has also led to the discovery of new types of variable white dwarfs, as the variable hot DQs, and the pulsating Extremely Low Mass white dwarfs. With the Kepler Mission, it has been possible to discover new phenomena, the outbursts present in a few pulsating white dwarfs.
}
\maketitle

\section{Introduction}\label{sec:intro}

White dwarf stars are the final evolutionary state of stars with initial masses up to 8.5--10.6~M$_\odot$\cite{Woosley15}, corresponding to 95 -- 97 \% of all stars.
The fraction
depends on the stellar metallicity, which affects both the Initial-Mass-Function and the Initial-to-Final-Mass Relation. 
For single stars, the minimum mass of a present day white dwarf is around 0.30--0.45~M$_\odot${}\cite{Kilic07}, 
because progenitors that would become lower mass white dwarfs have main sequence evolution time larger than the age of the Universe. 
Such masses correspond, considering the mass-radius relation of white dwarfs, to a minimal $\log g\simeq 6.5$. 
Evolutionary models e.g. by Ref.~\cite{Romero15} indicate that the maximum surface gravity for main sequence A stars, 
which have similar optical spectra to DA white dwarfs, corresponds to  $\log g \leq 4.75$, including very low metallicity.
There is therefore a gap between low mass white dwarfs and main sequence stars, $4.75~\leq~\log~g~\leq~6.5$.

Most white dwarfs do not generate energy from nuclear fusion, but radiate due to residual gravitational
contraction. Because of the degenerate equation of state, contraction is accompanied by a loss of thermal energy instead of increase
as in the case of ideal gases; the evolution of white dwarfs is therefore often simply described as cooling. The radius of an average white
dwarf star is of the same order of the Earth's radius, which implies that they have small surface area, resulting in very large
cooling times; it takes approximately $10^{10}$ years for the effective temperature of a $\sim 0.6 M_{odot}$ white dwarf to decrease
from $100\,000$~K to near $5\,000$~K. Consequently,
the cool $\sim 0.6 M_{odot}$ ones are still visible and among the oldest objects in the Galaxy\cite{GarciaBerro16}. Therefore, studying white dwarfs is extremely important to
comprehend the processes of stellar formation and evolution in the Milky Way\cite{Winget87,Campos16}. 

The progenitors of white dwarfs lose most of their envelope in the giant phases,
where mass loss depends on metallicity.
If the remainder H mass were above $\simeq 10^{-4} M*$, or the He mass above $\simeq 10^{-2} M*$ there would be observable nuclear burning in the white dwarf phase.  The limits depend on the mass of the white dwarf.
Most white dwarfs have atmospheres dominated by H, and the remainder by He. All other elements are only small traces, much less abundant than in the Sun, due to separation in the strong gravitational field\cite{Schatzman48}. 
The lightest elements float to the surface once the white dwarf cools below
The He-core white dwarf stars in the mass range $0.2-0.45~M_\odot$,
referred to as low-mass white dwarfs, are usually found in close binaries, often double degenerate systems\cite{Marsh95}, being most likely a product of interacting binary star evolution.
More than 70\% of those studied by Ref.~\cite{Kilic11} with masses below $0.45~M_\odot$ and all but a few with masses below $0.3~M_\odot$ show radial velocity variations\cite{Brown13,Gianninas14}.
Ref.~\cite{Kilic07} suggests single low-mass white dwarfs result from the evolution of old metal-rich stars that truncate
evolution before the helium flash due to severe mass loss. They also conclude all white
dwarfs with masses below $\simeq 0.3~M_\odot$ must be a product of binary
star evolution involving interaction between the components, otherwise
the lifetime of the progenitor on the main sequence would be
larger than the age of the Universe.

\begin{figure*}
\centerline{\includegraphics[width=0.9\textwidth]{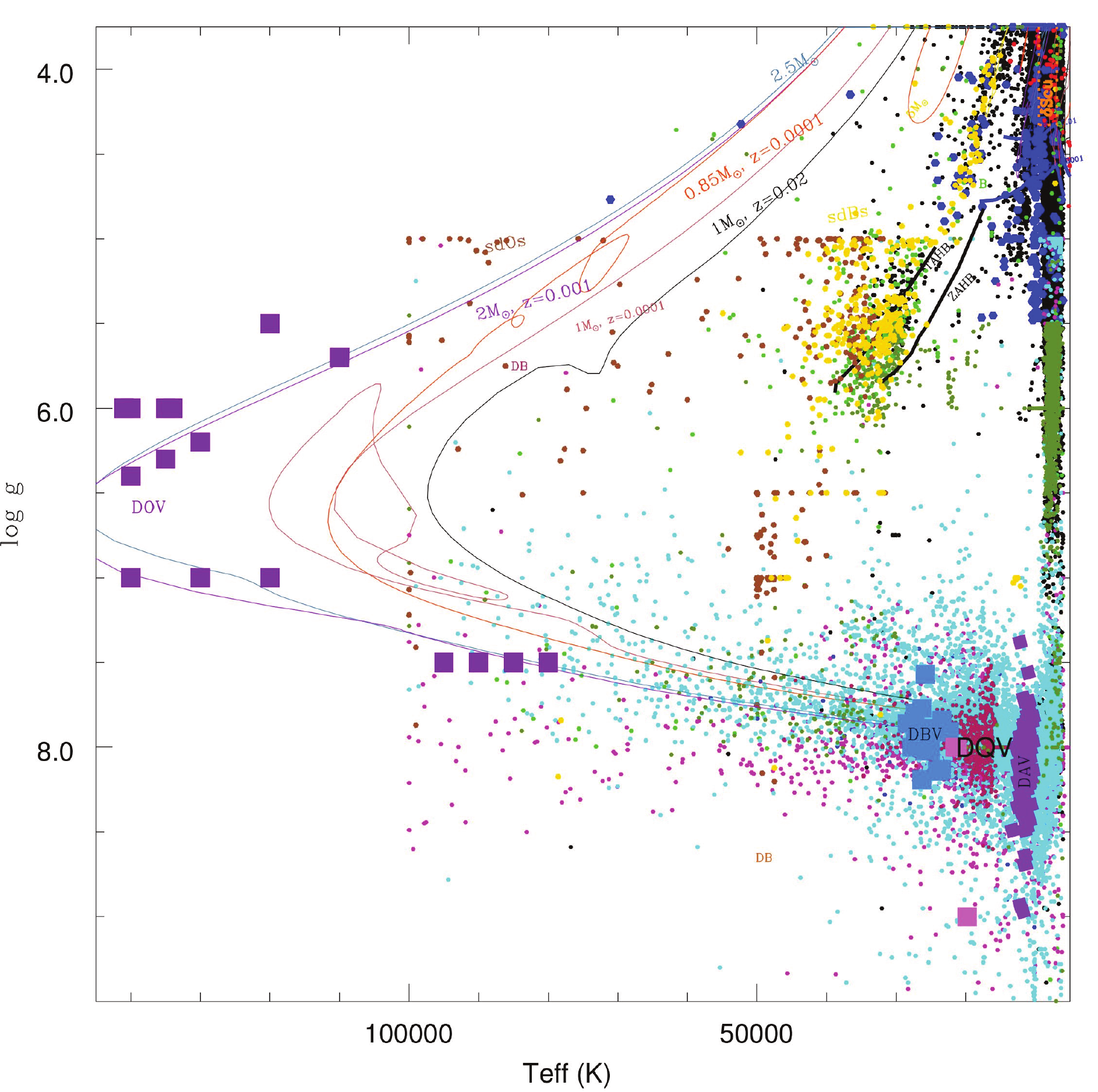}}
\vspace*{8pt}
\caption{Estimated effective temperature and superficial gravity for 79\,440 blue stars in the Sloan
Digital Sky Survey up to Data Release 13, and single star evolutionary models for different metallicities,
showing there is a region in the middle of the diagram that no single star evolutionary models cover.
Data of different colors represent distinct spectral classification and catalogs, DAs in light green, DBs in red, sdBs and sdOs in yellow, sdAs in dark green and black and the pulsating stars.
The Zero Age Horizontal Branch (ZAHB) and Terminal Age Horizontal Branch (TAHB) plotted were calculated with solar composition models.
\label{single}}
\end{figure*}

\begin{figure*}
\centerline{\includegraphics[width=0.9\textwidth]{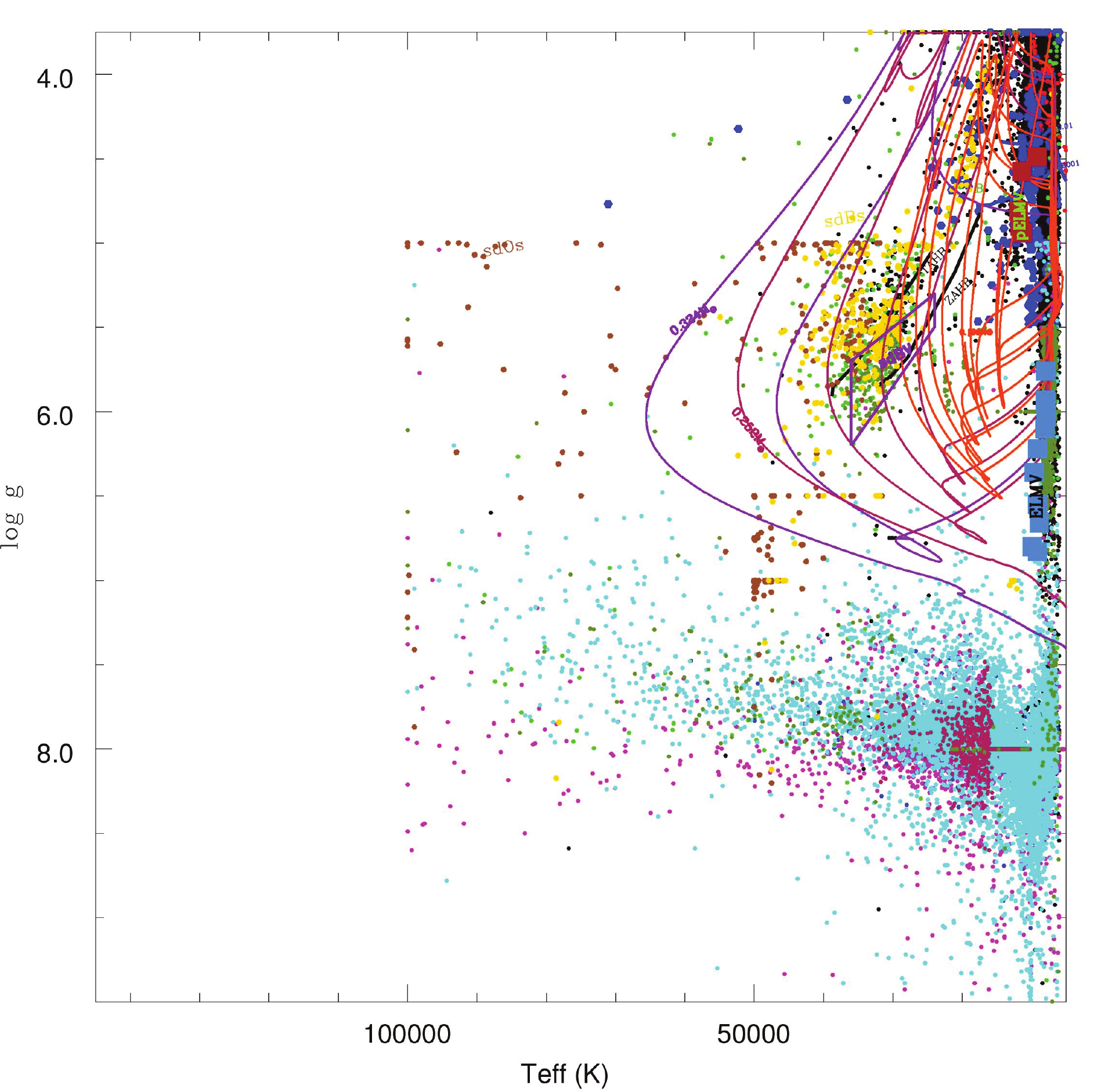}}
\vspace*{8pt}
\caption{Estimated effective temperature and superficial gravity for 79\,440 blue stars in SDSS Data Release 13, and interacting binary star evolutionary models\cite{Istrate16},
showing the region in the middle of the diagram is covered by interacting binary models. sdBs, SX~Phe, ELMs and pre-ELMs are in these regions.
\label{double}}
\end{figure*}

In Fig.~\ref{single} we show the results of our effective temperature and surface gravity determinations for all candidates from SDSS. We calculated the single star evolutionary models shown in the figure
with the MESA\cite{MESA} evolutionary code, including diffusion. In Fig.~\ref{double} the evolutionary models are those with rotation and diffusion of Ref.~\cite{Istrate16}.

Even though the low resolution hydrogen lines observed in SDSS spectra are poor surface gravity indicators below $T_\mathrm{eff} \simeq 10\,000$~K,
considering the SDSS spectra are concentrated mainly outside the Galaxy disk, we were surprised that several thousand stars were classified by the SDSS pipeline
as A and B stars. Considering their lifetimes on the main sequence smaller than 1~Gyr, and their distance modulus $(m-M)\geq 14.5$ at the SDSS bright saturation,
if these stars were main sequence stars, there would be a considerable population of young stars very far from the galactic disk. Using their measured radial
velocities, and proper motions if available, \cite{Pelisoli16} estimated their U, V, W velocities and show there would be a large number of hypervelocity A stars,
not detected up to date. If these stars are in fact low mass counterparts of interacting binary evolution, similar to the models of
\cite{Althaus13,Istrate16}, they are mainly concentrated in the galactic disk.
Considering we do not know their metallicities, and that low ionization potential metals contribute significantly to the electron pressure, we estimated their
surface gravities with two sets of models, a pure hydrogen model and a solar composition model. The estimated surface gravities with solar metallicity models
were on average $\Delta \log g \simeq 0.5$~dex smaller, but not systematically. Our plotted values are the solar metallicity ones.

\subsection{Interacting Binaries}
Ref.~\cite{Pietrzynski12} found an RR Lyrae with 0.26~$M_\odot$, and Ref.~\cite{Latour16} found a 0.23~$M_\odot$ pulsating subdwarf (sdBV).
Ref.~\cite{Istrate16} interacting binary with mass exchange models show that during a hydrogen shell burning pulse, an extremely low mass white dwarf
can cross the main sequence, horizontal branch and even giant instability strip. Ref.~\cite{Karczmarek17} estimate that up to 5\% of stars that cross the RR Lyrae and Cepheid instability strip are binaries.

DA white dwarf stars with masses $M\leq 0.45~M_\odot$
and $T_\mathrm{eff} < 20\,000$~K are Low Mass and Extremely Low Mass (ELM) as found by Refs.
\cite{Brown10}, \cite{Kilic11}, \cite{Brown12}, \cite{Brown13}, \cite{Gianninas14}, \cite{Gianninas15} and \cite{Brown16}.
Refs.~\cite{Hermes12} --
\cite{Bell16a} found pulsations
in eight of these ELMs,
similar to the pulsations seen in
DAVs (ZZ~Ceti stars), as described in Ref.~\cite{VanGrootel13}.
Ref.~\cite{Maxted14} found 17 pre-ELMs,
i.e., helium--core white dwarf precursors,
and
Ref.~\cite{Maxted14a,Gianninas16} report pulsations in six of them.
Pulsations are an important tool to study the stellar interior, and
Refs.~\cite{Corsico14} --
\cite{Istrate16a} report on theoretical models and pulsations of ELMs.
Refs.~\cite{Kepler16a} and \cite{Kepler16b}  show there are thousands of stars, photometrically classified as blue horizontal branch stars by Ref.~\cite{Xue08,Xue11,Carollo16}, that have spectroscopic
estimated surface gravities much higher than main sequence stars ($\log g \geq 4.75$) and therefore must have radii smaller than the Sun, classifying them as sdAs,
in line with the hot subdwarfs reviewed by Ref.~\cite{Heber16}.
Ref.~\cite{Pelisoli16}
discuss they are possibly Extremely Low Mass white dwarf stars.
Refs.~\cite{Kepler16a,Fusillo15} show that photometrically selected white dwarfs have a contamination around 40\%. Even the ones selected also from proper motion by Ref.~\cite{Munn17}
show significant contamination by non-white dwarf objects, when spectra are available.

Most stars that produced white dwarfs are born in binaries or multiple systems. Ref.~\cite{Lada06} demonstrates that while around 70\% of stars more massive than the Sun
are in binaries, two-thirds of the most common stars, M type dwarf stars, are single.
More than 10\% of the spectroscopically identified white dwarfs in SDSS have red companions\cite{Kepler16a,Rebassa16}.
Refs.~\cite{Farihi10,Nebot11} show that nearly 25\% of all main sequence binaries are close enough that mass transfer interactions occur when the more massive star becomes a red
giant or an asymptotic giant star.
If mass transfer exceeds the Eddington limit, the secondary star is not able to accrete the transferred material and the system evolves through a common envelope
phase, i.e., the core of the giant and the main sequence companion orbit within the
outer layers of the giant star, leading 
to the shrinkage of the orbit and the release of orbital energy. The orbital
energy deposited into the envelope eventually ejects it. Therefore a close binary is formed by the core
of the giant star and a main sequence companion, later a close white dwarf-main sequence binary. 
An ELM will be formed if the envelope is ejected before the helium-flash, which would happen if the star has a low initial mass, i.e., $M\lesssim 2 M_\odot$, 
to reach conditions to fuse helium in the core before it becomes degenerate.

\section{Mass Distribution}
We estimated the masses of all DA white dwarfs found by Ref.~\cite{Kleinman13}, \cite{Kepler15} and \cite{Kepler16a}. There were no new optical stellar spectra in SDSS Data Release 13. For the DA mass
distribution, we only consider spectra with S/N$\geq 15$ to have reliable mass determinations. 
From $T_\mathrm{eff}$ and  $\log g$ values obtained from our fits, after correcting to 3D convection following Ref.~\cite{Tremblay13a}, we use the
mass--radius relations of Refs.~\cite{Althaus05}, \cite{Renedo10} and \cite{Romero15}
to calculate the stellar mass.

Considering that white dwarfs with larger mass have smaller radius, and therefore can only be seen to smaller distances in a magnitude limited survey as SDSS, 
we calculated the density by correcting the visible volume with the $1/V_\mathrm{max}$ method of Ref.~\cite{Schmidt68},
up to a maximum g=19 magnitude.
For DAs with $T_\mathrm{eff} \geq 10000$~K, N=4054, we estimate $\langle M \rangle=0.647\pm 0.002~M_\odot$,
$T_\mathrm{eff} \geq 13000$~K, N=3637, $\langle M \rangle=0.646\pm 0.002~M_\odot$,
$T_\mathrm{eff} \geq 16000$~K, N=3012, $\langle M \rangle=0.641\pm 0.002~M_\odot$, 
$T_\mathrm{eff} \geq 25000$~K, N=1121, $\langle M \rangle=0.613\pm 0.003~M_\odot$.
The distribution shows that the DA and DB distributions have very different shapes. The DA's has a tail to larger masses, while the DB's is extended to lower masses. 
This is probably reflecting some limitation in the progenitors that can undergo very-late thermal pulses and become DBs.

\section{Pulsations}
During the cooling of the white dwarf star, partial ionization zones of C, O, He and H develop at subsequently lower $T_\mathrm{eff}$. Such partial
ionization zones increase the opacity and cause pulsations. For C/O, the stars are called pulsating PG~1159 stars, 
DOVs or GW~VIr stars, and occurs at 
$140\,000~\mathrm{K} \lesssim T_\mathrm{eff} \lesssim 75\,000$~K. 
For He, $32\,000~\mathrm{K} \lesssim T_\mathrm{eff} \lesssim 22\,000$~K and are called DBVs. For H,
$13\,000~\mathrm{K} \lesssim T_\mathrm{eff} \lesssim 10\,500$~K and are called DAVs or ZZ~Ceti stars.
Recent additions are the DQVs, with $22\,000~\mathrm{K} \lesssim T_\mathrm{eff} \lesssim 18\,000$~K, and
with $10\,000~\mathrm{K} \lesssim T_\mathrm{eff} \lesssim 8\,000$~K,
the ELMVs and the pre-ELMVs or EL CVn stars.

\begin{table}
\centering
\caption{Number of known pulsating white dwarfs.}
\begin{tabular}{lr}
\hline
Class & Number  \\\hline
DAVs & 181 \\   
DBVs & 23 \\
DOVs$^{1}$ & 22 \\
ELMVs & 11 \\
pre-ELMVs & 5 \\
DQVs & 3 \\\hline
\end{tabular}
\label{tab:tab-1}       
\\
\noindent {\footnotesize $^{1}$ Pulsating PG~1159 stars.}
\end{table}

\section{Magnetic Fields}
Ref.~\cite{GarciaBerro16a} presents a review on magnetic fields in white dwarf stars.
When examining each candidate SDSS spectrum by eye,
\cite{Kleinman13,Kepler15,Kepler16a}
found 822 stars with Zeeman splittings indicating magnetic fields
above 2~MG --- the limit where the line splitting becomes too small
to be identified at the SDSS spectral resolution \cite{Kepler13}.
The mean fields, estimated following \cite{Kulebi09}, range from 2~MG to 700~MG.
We caution that stars with large fields are difficult to identify because
fields above around 30~MG
intermixes subcomponents between different hydrogen series components so much that,
depending on effective temperature and signal-to-noise,
it becomes difficult to identify the star as containing hydrogen at all,
and affecting even the colors significantly.
Both the low field limit and the high
field limit are totally dominated by systematic effects, not the real limits. 
The effect of the magnetic field on pulsations has been estimated by
\cite[e.g.][]{Jones89,Tremblay15}.

\section{Rotation}
In general the measured rotation period for single white dwarfs ranges from 1~h to 18~d, with a median around 1~d\cite{Kawaler15}.
The fastest single white dwarf rotator from asteroseismological measurements (Table~\ref{rot}) is
the $0.79~M_\odot$ DAV SDSS~J161218.08+083028.1 discovered by
Ref.~\cite{Castanheira13}, assuming the two observed periods at 115.0~s and 117.0~s are two components of a rotation triplet.

\begin{table}[!ht]
\caption{Rotation periods of white dwarfs as determined via asteroseismology.}
{\begin{tabular}{lcccc}
\noalign{\smallskip}
Star & $P_{\rm rot}$ [h]  & $T_\mathrm{eff}$\footnote{The effective temperatures and masses are corrected to 3D convection\cite{Tremblay13}.} & Type & $M$ [$M_{\odot}$] \\
\noalign{\smallskip}
RX J2117.1+3412  & 28  & 170000 & GW Vir & 0.72 \\
PG 1159-035   & 33  & 140000 & GW Vir & 0.54 \\
NGC 1501  & 28  & 134000 & [WCE] & 0.56 \\
PG 2131+066   &  5  &95000 & GW Vir & 0.55 \\
PG 1707+427   & 16  & 85000 & GW Vir & 0.53 \\
PG 0122+200   & 37  & 80000 & GW Vir & 0.53 \\

PG 0112+104 & 10.17 & 31040 & DBV & 0.58 \\
KIC 8626021 & 43  & 29700 & DBV & 0.56 \\
EC 20058-5234 &  2  & 25500 & DBV & 0.65 \\
GD 358    &   29  & 23740 & DBV & 0.54 \\

SDSS J083702.16+185613.4 & 1.13 & 13590 & ZZ Ceti & 0.88 \\ 
G 226-29                 &  9 & 12510 & ZZ Ceti & 0.83 \\
G 185-32                 & 15 & 12470 & ZZ Ceti & 0.67 \\
SDSS J113655.17+040952.6 & 2.6 & 12330 & ZZ Ceti & 0.55 \\
SDSS J161218.08+083028.1 & {\bf 0.93} & 12330 & ZZ Ceti & 0.79 \\
Ross 548                 & 37 & 12300 & ZZ Ceti & 0.63 \\
GD 165                   & 50 & 12220 & ZZ Ceti & 0.68 \\
LP 133-144               & 41.8 & 12150 &ZZ Ceti & 0.59 \\
KIC 11911480             & 86.4 & 12160 &ZZ Ceti & 0.58 \\
L 19-2                   & 13 & 12070 & ZZ Ceti & 0.69 \\
HS 0507+0435             & 41   &  12010 & ZZ Ceti & 0.73 \\
EC 14012-1446            & 14.4 & 12020 & ZZ Ceti & 0.72 \\  
KUV 11370+4222           & 5.56 &  11940 & ZZ Ceti & 0.72 \\
G 29-38                  & 32 &   11910 & ZZ Ceti & 0.72 \\
KUV 02464+3239           & 90.7 & 11620 & ZZ Ceti & 0.70 \\  
HL Tau 76                & 53   &  11470 & ZZ Ceti & 0.55 \\
SDSS J171113.01+654158.3 & 16.4 &  11130 & ZZ Ceti & 0.90 \\
GD 154                   & 50.4 &  11120 & ZZ Ceti & 0.65 \\
KIC 4552982              & 15.0 &  10860 & ZZ Ceti & 0.71 \\
SDSS J094000.27+005207.1 & 11.8 & 10590 & ZZ Ceti & 0.82 \\
\end{tabular}\label{rot}}
\end{table}

Differential rotation in white dwarfs was studied by
Refs.  \cite{Charpinet09} -- 
\cite{Hermes16}, 
using the change in rotation splitting of non-radial pulsations.

\section{Axions and Dark Mass}
Axions are the best candidates for dark mass\cite{Ringwald16}. Refs.\cite{Isern03,Isern10,Corsico12,Corsico12a,Corsico16,Battich16} show white dwarf pulsations and luminosity function are
consistent with extra cooling caused by axions of masses around $17\pm 4$~meV.

\section{Fitting Models}
The pulsation spectra exhibited by ZZ Ceti stars strongly depends on their inner chemical profile. 
There are several processes affecting the chemical profiles that are still not accurately determined.
See \cite{Geronimo17} for a study of the impact of the current uncertainties in stellar evolution on the expected pulsation properties of ZZ Ceti stars.

Each single period is determined by the integral of the pulsation kernel (or work function) over
the whole star. It cannot distinguish among different distributions, as demonstrated for
example by \cite{Montgomery03}.
If different modes are not independent, i.e., they sample the same regions of the distributions, 
they carry less information than independent modes.
\cite{Giammichele16} propose one could determine the whole chemical distribution profile from pulsations. 
\cite{Giammichele17} performed a test using ten periods, namely the modes l=1, k=2,3,4,5,6 and l=2, k=3,4,5,6,7. 
Note that in this tests, a sequence of consecutive modes k=2,3,4,5,6 is needed to sample the structure, 
sequence that is usually not seen real stars. 
ZZ Ceti stars shows different pulsation spectra: hot stars show few short modes that sample the
inner parts of the star while the  cool stars show many long modes, which sample the outer parts of the stars. 
These characteristics must be taken into account when asteroseismology is applied to white dwarfs. 
A good example of a hot ZZ Ceti pulsator is G~117-B15A, with three modes. Were it not for the convection 
description problem, that introduces an uncertainty of 
$\Delta T_{\rm eff} \simeq 500$~K,
and the problem with line broadening that gives different  $\log g$ from different spectral lines, 
we could use the 3 modes of G~117-B15A to get 3 structural parameters, plus $dP/dt$ to estimate the core mean molecular weight.

\section{Kepler satellite}
Observations of pulsating white dwarfs with the Kepler satellite are limited to the brightest objects due to the relatively small
size of the telescope, but its long observations has allowed not only an exquisite precision in the pulsation spectra but also
the discovery of outbursts lasting hours (\cite{Bell15,Hermes15,Bell16}). 
These outbursts resemble the {\it forte} episode observed in 1996 for the
DBV GD~358 by \cite{Nitta98,Kepler03}.

\begin{acknowledgement} 
\noindent\vskip 0.2cm
\noindent {\em Acknowledgments}: S.O.K. and A.D.R. are financed by Conselho Nacional de Desenvolvimento
Cient\'{\i}fico e Tecnol\'ogico, Brasil.
This research has made use of NASA's Astrophysics Data System and of the cross-match service provided by CDS, Strasbourg. 
Funding for the Sloan Digital Sky Survey has been provided by the Alfred P. Sloan Foundation, the U.S. Department of Energy Office of Science, and the Participating Institutions. The SDSS web site is www.sdss.org.
Part of the work has been based on observations obtained at the Southern Astrophysical Research (SOAR) telescope, which is a joint project of the Minist\'erio da Ci\^encia, Tecnologia, Inova\c{c}\~oes e Comunica\c{c}\~oes (MCTIC) da Rep\'ublica Federativa do Brasil, the U.S. National Optical Astronomy Observatory (NOAO), the University of North Carolina at Chapel Hill (UNC), and Michigan State University (MSU).
\end{acknowledgement}

\end{document}